\documentclass[doublecol]{epl2}
\usepackage{amsmath}

\title{Underdamped Quantum Ratchets}
\shorttitle{Underdamped Quantum Ratchets}

\author{Sergey Denisov \and Sigmund Kohler \and Peter H\"{a}nggi}
\shortauthor{S. Denisov \etal}

\institute{Institut f\"ur Physik, Universit\"at  Augsburg,
       Universit\"atsstra{\ss}e~1, D-86135 Augsburg, Germany
}

\abstract{We investigate the quantum ratchet effect under the
influence of weak dissipation which we treat within a Floquet-Markov
master equation approach. A ratchet current emerges when all
relevant symmetries are violated.  Using time-reversal symmetric
driving we predict a purely dissipation-induced quantum ratchet
current. This directed quantum transport results from bath-induced
superpositions of non-transporting Floquet states.}

\begin{document}

\maketitle

An intriguing phenomenon in non-equilibrium transport is
the ratchet effect \cite{Smol, Han1, Reim}, \textit{i.e.}, the emergence of
directed motion in the absence of any net bias. Net transport
results from an interplay between ac driving, spatio-temporal
asymmetries, and non-linearities in a periodic potential.  This
mechanism provides the basis for an increasing number of experiments
ranging from particle transport in biological systems \cite{biolog} and
nano-engines \cite{HH} to charge transport in semiconductor
heterostructures \cite{Linke1999a, Khrapai2006a}, superconductors
\cite{Majer2003} and spin transport \cite{Grifoni_spin}. Symmetry
investigations revealed the necessary conditions on the ac force and
the static potential, such that a ratchet current can emerge
\cite{Flach1, Kohler1, Den, ren}.

A widely employed model for studying the ratchet effect is a
one-dimensional periodic potential in which classical Brownian
particles move \cite{Smol, Han1, biolog, Reim, HH}.  It describes also the
motion of a thermal cloud of cold atoms in an ac-driven optical
potential \cite{ren}. As the atom cloud is cooled down further, one
expects quantum effects to become relevant \cite{very_cold}.  The
Hamiltonian limit of such quantum ratchets has been studied recently
\cite{Ketzmerick, Casati0, ac-quantum}.
A more realistic description of quantum ratchets necessitates
inclusion of the ubiquitous decoherence and quantum dissipation
\cite{Han_quantum, Goychuk, Grifoni_diss, Scheidl}. For
moderate-to-strong dissipation, incoherent tunneling transitions
prevail and the quantum ratchet current can be studied within quantum
rate theory \cite{Han_quantum, Goychuk, Grifoni_diss, Scheidl}, while
in the high-temperature limit, one can employ a Fokker-Planck equation
with quantum corrections \cite{Zueco}.  For
very strong friction, a description in terms of an effective
Smoluchowski equation comprising leading-order quantum corrections is
appropriate \cite{Han_Sm}. By contrast, the crossover towards the
coherent quantum regime, \textit{i.e.}, the underdamped regime \cite{Casati},
in which already weak decoherence significantly alters the Hamiltonian
dynamics, still represents an ambitious challenge.

In this letter we study ac-driven quantum ratchet transport in the
technically demanding regime of weak quantum dissipation where
quantum coherence and relaxation affect each other. We analyze
within a Floquet-Markov description \cite{Kohler} the dynamics on
quantum attractors by expanding them into the Floquet states of the
corresponding coherent time-dependent system. Then a most intriguing
question is whether violation of time-reversal symmetry due to weak
quantum dissipation is perceivable in the quantum attractor and in
the quantum ratchet current.

\section{Model and quantum master equation}

A quantum particle in a time-dependent periodic potential
obeys the Schr\"{o}dinger equation
\begin{eqnarray}
\label{Ham1} \mathrm{i} \hbar \frac{\partial}{\partial t}{\psi}(x,t) &=&
\Big[-\frac{\hbar^{2}}{2m}\frac{\partial^{2}}{\partial
x^{2}}+V(x,t)\Big]
     \psi(x,t),
\\
V(x,t) &=& V_{0} u(x)-x E(t),\nonumber
\end{eqnarray}
where $u(x)=u(x+L)$ with $\mathop{\mathrm{max}}|u(x)| \sim 1$
describes the shape of the $L$-periodic potential. The driving
$E(t)$ is a time-periodic field with zero mean,
$E(t+\mathcal{T})=E(t)$, $\langle E(t) \rangle_{\mathcal{T}}=0$.
Henceforth, we use $1/k_L\equiv L/2\pi$, $(m/k_L^2V_0)^{1/2}$, and
$V_0$ as units of distance, time, and energy, respectively, such
that formally $k_L=m=V_0=1$, while $\hbar$ becomes the effective
Planck constant $\hbar k_L/\sqrt{m V_0}$ \cite{very_cold}.

By the gauge transformation $|\psi \rangle \rightarrow
\exp(-\frac{\mathrm{i}}{\hbar} x A(t)) |\psi \rangle$, we bring the
Schr\"odinger equation (\ref{Ham1}) to the spatially periodic form
\cite{ac-quantum}
\begin{equation}
\mathrm{i} \hbar \frac{\partial}{\partial t}{\psi}(x,t)
=\Big(\frac{1}{2}[\hat{p}-A(t)]^{2}+u(x)\Big)\psi(x,t), \label{Ham2}
\end{equation}
with the vector potential $A(t)=-\int_{0}^{t}E(t')dt'$ and the
momentum operator $\hat p = -\mathrm{i}\hbar\partial/\partial x$.  The
corresponding Hamiltonian has recently been realized in cold atom
experiments \cite{very_cold}. The Schr\"odinger equation
(\ref{Ham2}) is time periodic with period $\mathcal{T}=2 \pi/\omega$
and, thus, according to the Floquet theorem, it possesses a complete
set of mutually orthogonal solutions of the form $
|\psi_{\alpha}(t)\rangle= \mathrm{e}^{-\mathrm{i}{\epsilon_{\alpha}}t/{\hbar}}
|\phi_{\alpha}(t)\rangle$.  The Floquet states
$|\phi_{\alpha}(t)\rangle=|\phi_{\alpha}(t+\mathcal{T})\rangle$ and
the quasienergies $\epsilon_{\alpha}$, $-\hbar \omega/2  <
\epsilon_{\alpha} < \hbar \omega/2$, are obtained from the
eigenvalue problem $[H(t)-\mathrm{i}\hbar\partial/\partial t]|\phi(t)\rangle
=\epsilon|\phi(t)\rangle$ \cite{Rotating}. Owing to discrete
translation invariance, all Floquet states are characterized by a
quasi momentum $\kappa$ with $|\phi_{\alpha}(x+2\pi)\rangle =
\exp(\mathrm{i} \hbar \kappa) |\phi_{\alpha}(x)\rangle$. We restricted
our study to states with $\kappa=0$, which can be expanded into
the plane waves $|n\rangle = (2\pi)^{-1/2}\exp(\mathrm{i}nx)$.
Physically, these states correspond to initial states with atoms
populating all wells of the spatial potential equally.
Such initial conditions are the natural ones for the recently
proposed \cite{amico} ring-shaped optical potentials which have
already been realized experimentally \cite{ring_exp}.

We incorporate decoherence and dissipation by coupling the driven system
(\ref{Ham2}) to a bath of non-interacting harmonic oscillators
\cite{Caldeira}. Following a standard approach to weak quantum
dissipation \cite{Kohler}, we decompose  the reduced density
operator $\varrho$ into the Floquet basis of the coherent system,
$\varrho_{\alpha \beta}(t)=\langle \phi_{\alpha}(t) | \varrho(t) |
\phi_{\beta}(t) \rangle$.
Assuming that dissipative effects are relevant only on time scales
much larger than the driving period $\mathcal{T}$, we arrive at the
master equation
\begin{eqnarray}
\dot{\varrho}_{\alpha\beta} =
-\frac{\mathrm{i}}{\hbar}(\epsilon_{\alpha}-\epsilon_{\beta})
  \varrho_{\alpha\beta} +\sum_{\alpha' \beta'}
  \mathcal{L}_{\alpha \beta, \alpha'\beta'}\,\varrho_{\alpha'\beta'},
 \label{full}
\end{eqnarray}
with the time-independent transition rates
\begin{equation}
\begin{split}
\mathcal{L}_{\alpha \beta, \alpha' \beta'}
= & \sum_{n}(N_{\alpha \alpha', n} + N_{\beta \beta', n})
    X_{\alpha \alpha', n} X_{\beta \beta', -n}
\\& -\delta_{\beta \beta'}\sum_{\beta'', n} N_{\beta'' \alpha', n}
    X_{\alpha \beta'',-n} X_{\beta'' \alpha', n}
\\& -\delta_{\alpha \alpha'}\sum_{\alpha'', n} N_{\alpha'' \beta', n}
    X_{\beta' \alpha'',-n} X_{\alpha'' \beta, n} \,,
\end{split}
\label{rates}
\end{equation}
where $X_{\alpha\beta,n} = \langle\langle\phi_\alpha(t)|x\,
\mathrm{e}^{-\mathrm{i}n\omega t}
|\phi_\beta(t)\rangle\rangle_{\mathcal{T}}$ and $N_{\alpha\beta,n} =
N(\epsilon_\alpha-\epsilon_\beta+n\hbar\omega)$
with $N(\epsilon) = ({\gamma\epsilon}/{\hbar^2}) n_{\rm
th}(\epsilon)$. Here $n_{\rm th}(\epsilon)=[\exp({\epsilon/k_{\rm
B}T}) - 1]^{-1}$ is the thermal occupation number and $\langle\cdots
\rangle_{\mathcal{T}}$ denotes the average over one driving period.

This Markov approximation requires that the coupling strength
$\gamma$ is the smallest frequency scale in the problem, such that
$\gamma \ll k_{\rm B}T/\hbar$ and $\gamma \ll
\Delta_{\alpha\beta}/\hbar$, where
$\Delta_{\alpha\beta}=|\epsilon_{\alpha}-\epsilon_{\beta}|$ is any
splitting of the Floquet spectrum \cite{Kohler}.
The latter condition is rather strict because in the present case,
the quasienergies are even dense on the interval $[-\hbar\omega/2,
\hbar\omega/2]$.  Thus, the condition $\gamma \ll
\Delta_{\alpha\beta}/\hbar$ is violated for any finite dissipation
strength.  Yet it is obvious that only a finite number of Floquet
states significantly contributes to the density matrix at large
times.  We thus validate our results for the asymptotic state
$\varrho_{\alpha\beta}$ with the following criterion: We sort the
Floquet states according to their weights $\varrho_{\alpha \alpha}$
and consider only results for which the first $N_\varepsilon$ states
fulfill the condition $\gamma/\Delta_{\alpha\beta} < \varepsilon$,
where $\varepsilon$ is the threshold value.

For any initial density operator $\varrho_{\alpha \beta}(0)$, the
solution of the master equation~(\ref{full}) converges to a unique
``quantum attractor'' being the fixed point
$\varrho_{\alpha\beta}^{a} = \varrho_{\alpha \beta}(t \rightarrow
\infty)$ of the quantum master equation. Note that in the
Schr\"odinger picture, the density operator is periodically
time-dependent and, thus, describes a limit cycle.  Since in Floquet
representation, the attractor $\varrho^{a}_{\alpha \beta}$ is
nevertheless time-independent, the asymptotic current, defined as
the time-averaged momentum expectation value, reads
\begin{equation}
J = \sum_{\alpha \beta} \varrho^{a}_{\alpha \beta} \bar
p_{\alpha\beta}; \quad \bar p_{\alpha\beta} = \langle\langle
\phi_{\alpha}(t)| \hat{p}| \phi_{\beta}(t)\rangle
  \rangle_\mathcal{T}.
\label{Current2}
\end{equation}

A frequently used simplification is possible if dissipative effects
are relevant only on time scales longer than any $2 \pi
\hbar/(\epsilon_{\alpha}-\epsilon_{\beta}
-\epsilon_{\alpha'}+\epsilon_{\beta'})$.  Then, one can employ a
full rotating-wave approximation (RWA) for which the diagonal and
off-diagonal density matrix elements decouple, such that the quantum
attractor becomes diagonal, \textit{\textit{i.e.}}, $\varrho^a_{\alpha\beta} =
\varrho^a_{\alpha\alpha} \delta_{\alpha\beta}$.

\section{Symmetries}

Before evaluating the ratchet current we determine two symmetry
conditions under which the current vanishes.
The first one is the generalized parity $S:(x,t) \to
(-x,t+\mathcal{T}/2)$ \cite{Rotating}, which is present if the
potential and the driving field fulfill the relations
\begin{equation}
u(-x)= u(x); \quad E(t+\mathcal{T}/2) = -E(t). \label{eq:Ta}
\end{equation}
Then, the Floquet states obey $\phi_{\alpha}(-x,t+\mathcal{T}/2)
=\sigma_{\alpha} \phi_{\alpha}(x,t)$, where $\sigma_{\alpha}=\pm 1$
according to the generalized parity.  As a consequence, we find $\bar
p_{\alpha\beta} = -\sigma_\alpha\sigma_\beta \bar p_{\alpha\beta}$
and thus in particular $\bar p_{\alpha\alpha}=0$.  This means that all
Floquet states are non-transporting on time-average \cite{ac-quantum},
such that any non-vanishing current must stem from off-diagonal
density matrix elements.
The master equation (\ref{full}) inherits a symmetry from the
position matrix elements for which the relation $X_{\alpha \beta n}
= (-1)^{n+1}\sigma_{\alpha}\sigma_{\beta} X_{\alpha \beta n}$ holds
\cite{Kohler1}. This leads to the conclusion that the asymptotic
state obeys $\varrho^a_{\alpha\beta} =
\sigma_\alpha\sigma_\beta\varrho^a_{\alpha\beta}$. Inserting these
symmetry relations into expression~(\ref{Current2}) yields $J = -J$,
which implies that the ratchet current vanishes in the presence of
generalized parity.

A second relevant symmetry is time-reversal symmetry $t\to-t$.  It
has the consequence that if $\varrho(t)$ is a solution of the master
equation (\ref{full}), then $\varrho(-t)$ is a solution as well.
Time-reversal symmetry is present if the driving field obeys
\begin{equation}
E(t+t_\mathrm{s})=E(-t+t_\mathrm{s}), \label{eq:Tb}
\end{equation}
with some appropriate time shift $t_\mathrm{s}$, and obviously can
persist only in the Hamiltonian limit $\gamma=0$, for which the
dynamics depends on the initial conditions.  In the Hamiltonian limit,
a meaningful ratchet current requires averaging over all possible
initial conditions.

Let us again emphasize that all in the presence of either of the two
symmetries in eqs.~(\ref{eq:Ta}), (\ref{eq:Tb}), Floquet states are
non-transporting such that their current $\bar p_{\alpha\alpha}$
vanishes, cf.\ Ref.~\cite{ac-quantum}.  Since within full RWA by
construction $\varrho^{a}_{\alpha\beta} = 0$ for $\alpha\neq\beta$,
while $\bar{p}_{\alpha\alpha} = 0$, the current (\ref{Current2})
vanishes within this approximation as well.  This in turn means that
the purely dissipation-induced ratchet current studied below can be
obtained only from the full master equation (\ref{full}).

\begin{figure}[t]
\begin{center}
\includegraphics[height=3.8cm]{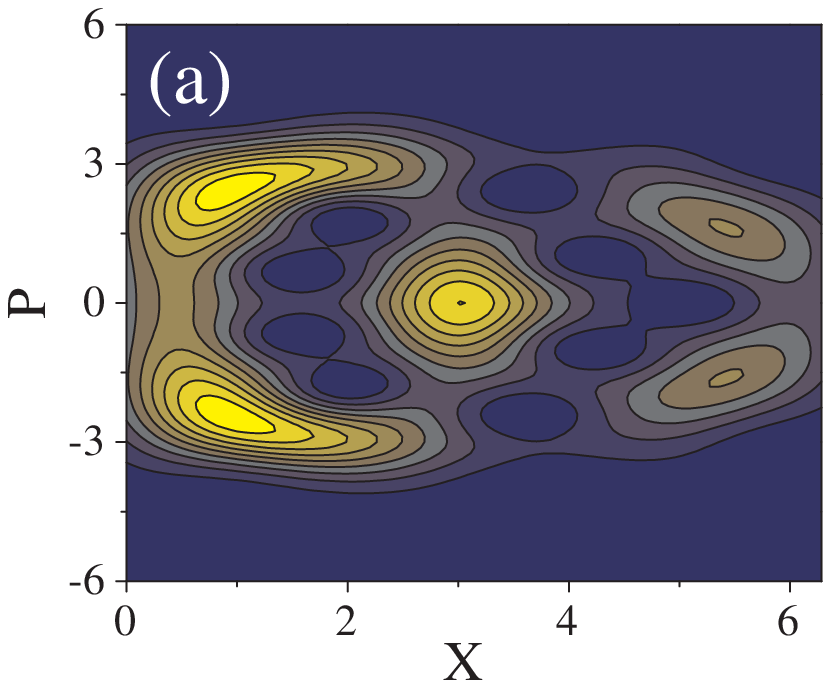}
\includegraphics[height=3.8cm]{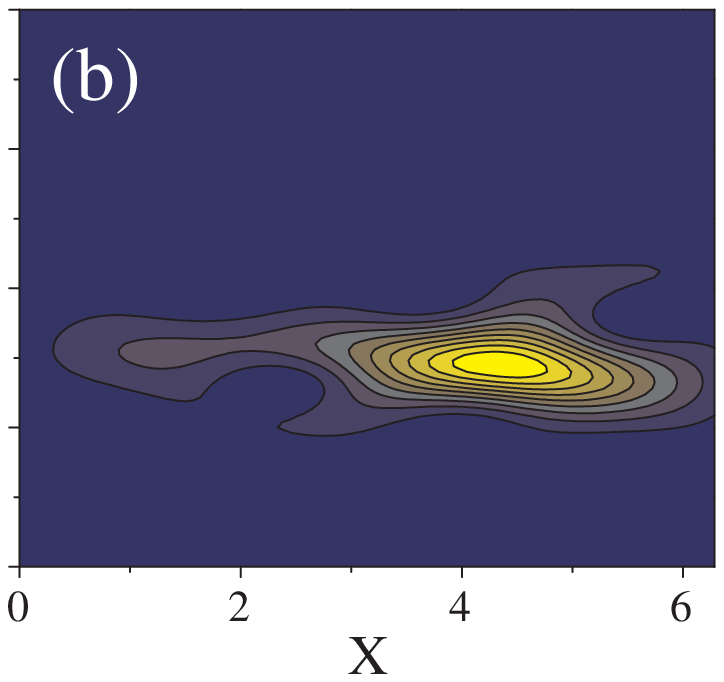}
\end{center}
\caption{(Color online) Husimi representation of the eigenstate
for the potential~(\ref{driv}) for phase lag $\theta=0$ (a) and
$\theta=-\pi/2$ (b). The corresponding momenta are
$\bar{p}_{\alpha\alpha}=0$ (a) and $\bar{p}_{\alpha\alpha}\approx
-0.15$ (b) (in units of the recoil momentum). The eigenstate has been
tracked along the corresponding quasienergy band. The parameters are
$\hbar=0.5$, $E_1=1.6$, $E_2=2$, $\omega=1$.}
\label{Fig:Floquet}
\end{figure}

In order to observe a quantum ratchet current, we need to specify
the periodic potential $u(x)$ and the driving field $E(t)$ such that
at least one of the conditions in eq.~(\ref{eq:Ta}) is violated.%
\footnote{For the time evolution of a localized atom cloud in an
extended periodic optical potential \cite{ren, Weitz2007}, one must
consider the whole range of quasimomenta, $\kappa \in [-\pi, \pi]$
\cite{Ketzmerick}. Then the symmetry transformations involve
eigenstates with opposite $\kappa$. Neverhteless, the skewed initial
conditions generally yield a nonzero current even in symmetric
potentials.}
One possibility would be to use a non-reflection-symmetric static
potential $u(x)$ together with a sinusoidal driving
\cite{Weitz2007}. Here, by contrast, we consider a
\textit{symmetric} potential and a bichromatic driving field,
\textit{i.e.},
\begin{equation}
u(x) = \cos(x);\quad E(t)=E_{1}\cos (\omega t)+E_{2}\cos (2\omega t
+\theta) , \label{driv}
\end{equation}
which breaks generalized parity provided that both $E_1$ and $E_2$
are non-zero \cite{ac-quantum}.  Indeed, if either $E_1=0$ or
$E_2=0$, the ratchet current vanishes, both classically and quantum
mechanically.  The phase lag $\theta$ allows one to control the
time-reversal symmetry: If $\theta$ is a multiple of $\pi$, the
driving field obeys the symmetry condition (\ref{eq:Tb}).  For any
other phase lag and $E_1,E_2\neq 0$, time-reversal symmetry is
broken, as can be seen in the Husimi functions \cite{Husimi} of the
Floquet states depicted in fig.~\ref{Fig:Floquet}.
Moreover, since the transformation $(x,\theta) \rightarrow
(-x,\theta \pm \pi)$ leaves the Hamiltonian (\ref{Ham2}) invariant
while it inverts the current, we find that in the Hamiltonian limit,
the ratchet current obeys \cite{ac-quantum}
\begin{eqnarray}
J(\theta)=-J(\theta+\pi) =-J(-\theta). \label{SymSym}
\end{eqnarray}
Notably, this relation does not hold for finite dissipation strength
$\gamma>0$.
\begin{figure}[t]
\centerline{\includegraphics[width=6.5cm]{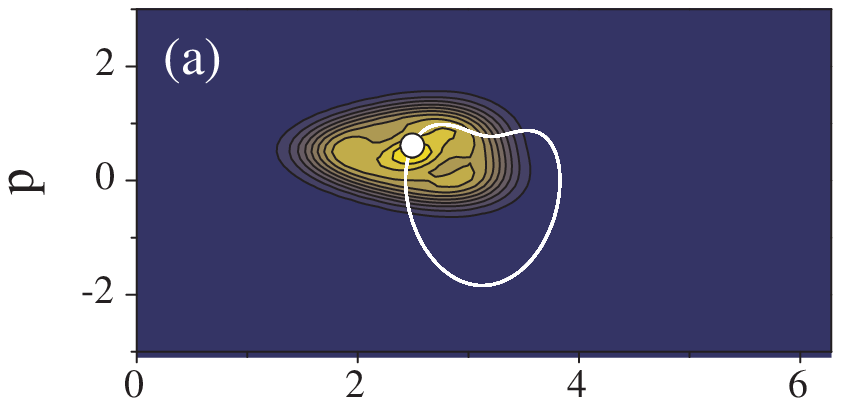}}
\centerline{\includegraphics[width=6.5cm]{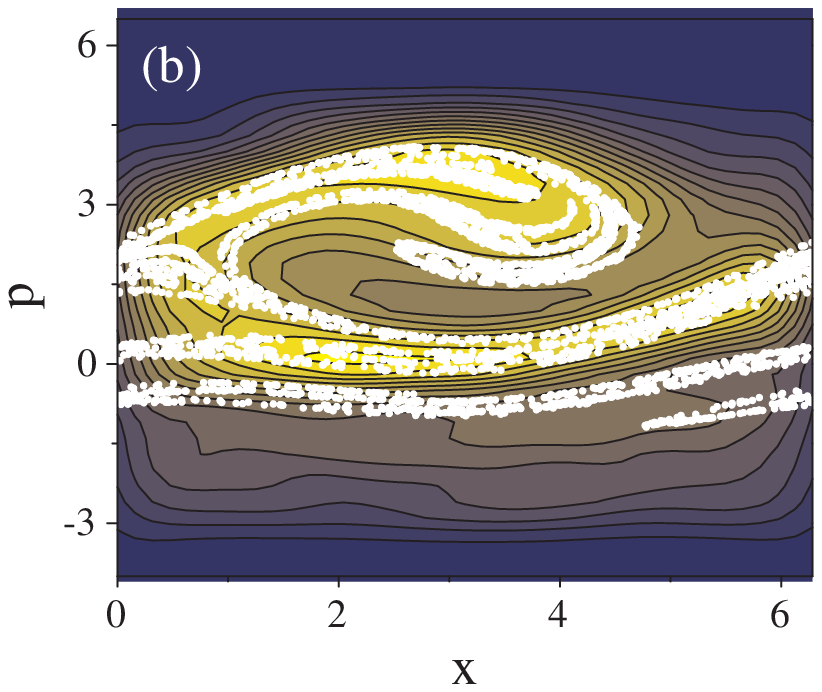}}
\caption{(Color online) Husimi representation of the quantum
attractor at stroboscopic times with the corresponding classical
attractor superimposed (white dots) for $E_1=E_2=2$, $\gamma=0.1$,
$k_\mathrm{B}T=0.1$.  Driving frequency, phase lag, and effective
Planck constant are $\omega=2$, $\theta=\pi/2$, $\hbar=0.1$ (a) and
$\omega=0.87$, $\theta=-\pi/2$, $\hbar=0.2$ (b). The white line in
panel (a) marks the corresponding limit cycle. The numerical
integration has been performed with 45 basis states and $\varepsilon
= 0.3$.} \label{Fig:husi}
\end{figure}%

\section{Quantum ratchet current and quantum attractor}

We start out our numerical studies of the dissipative quantum system
by validating our master equation approach for the quantum-classical
correspondence.
Already a classical Brownian particle in the driven periodic
potential~(\ref{driv}) exhibits a rather rich dynamics, ranging from
regular limit cycles to chaotic motion on strange attractors, see
fig.~\ref{Fig:husi}.  The corresponding quantum dynamics is even
more complex: In the deep quantum regime $\hbar \geq 1$, it is
restricted to a few Floquet states.  In the semiclassical regime
$\hbar\ll 1$, by contrast, many levels play a role and, thus, we
expect to find in the Husimi representation of the density operator
signatures of the classical phase-space structure.  This represents
a demanding requirement for our master equation formalism.  In order
to emphasize the power of the Floquet master equation (\ref{full}),
we plotted the Husimi function of the quantum attractor for both a
regular limit cycle [fig.~\ref{Fig:husi}(a)] and a strange attractor
[fig.~\ref{Fig:husi}(b)].  Comparison with the corresponding
classical attractors underlines that our formalism is able to cope
with the semiclassical limit.%
\footnote{The classical dissipative equations of motion corresponding
to the Hamiltonian (\ref{Ham2}) read $\dot x=p$, $\dot{p}=-\gamma p +
\sin(x) + E(t)+\gamma A(t)$.}

Since the classical attractor shown in fig.~\ref{Fig:husi}(a) is
bounded, it is non-transporting, $J_\mathrm{cl}=0$.
The quantum attractor, by contrast, supports a very small, but
finite dc current, $J_\mathrm{qm} \simeq 0.0025$. Note that the dc
current for the chaotic attractor shown in fig.~\ref{Fig:husi}(b)
is much larger, namely $J_\mathrm{cl} = 0.45$ and
$J_\mathrm{qm}=0.32$, respectively.
\begin{figure}[t]
\begin{center}
\includegraphics[width=0.85\linewidth,clip=true]{fig3a.eps}
\\
\includegraphics[height=3.7cm]{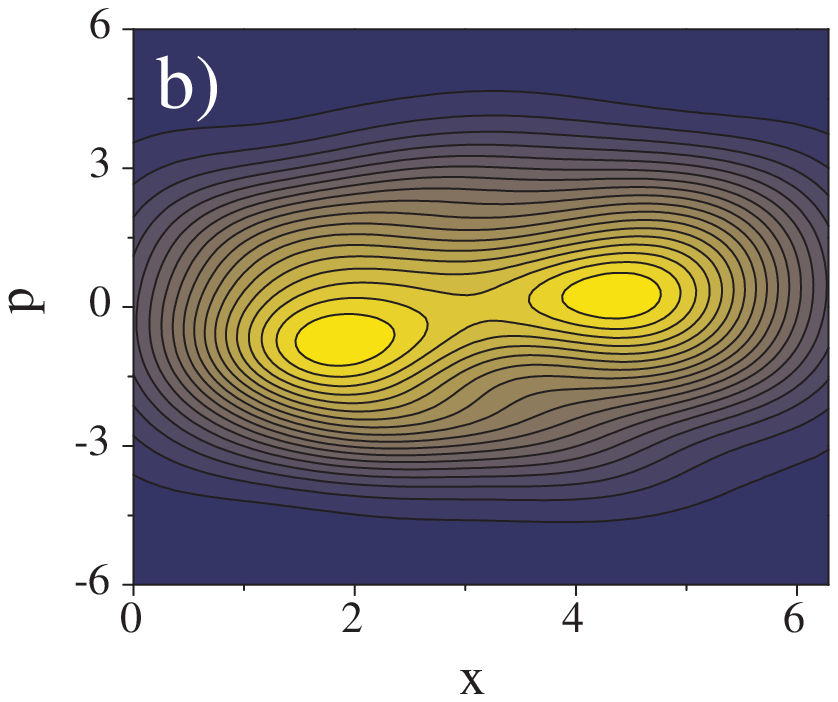}
\includegraphics[height=3.7cm]{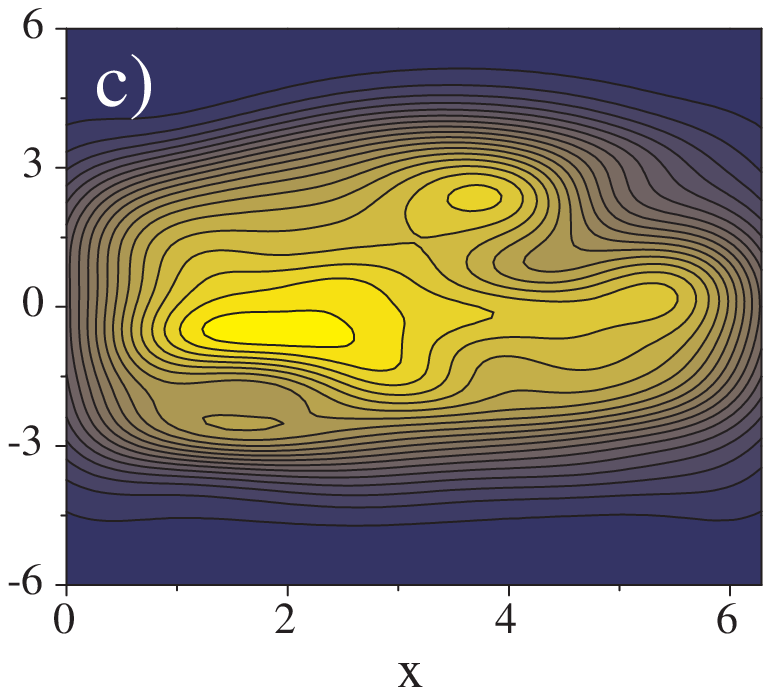}
\end{center}
\caption{(Color online) (a) Average current $J$ (in units of the
recoil momentum) as a function of the phase lag $\theta$ for
dissipation strength $\gamma=0.02$ (solid line) and
$\gamma=5\times10^{-4}$ (dashed) compared to the Hamiltonian case
(dash-dotted). The inset shows the current for $\theta = 0$ as a
function of the dissipation strength $\gamma$.
Lower panel: Husimi representation of quantum attractors at
stroboscopic times $t=nT$ for $\theta=0$: (b) $\gamma=5\cdot 10^{-4}$
and (c) $\gamma=0.02$. The parameters are $\hbar=1$, $E_1=1.6$,
$E_2=2$, $\omega = 1$, and $k_\mathrm{B}T=0.5$.  The numerical
integration has been performed with 21 basis states and $\varepsilon
= 0.3$} \label{Fig:current}
\end{figure}%

Figure \ref{Fig:current}(a) depicts the ratchet current as a function
of the phase lag $\theta$ for different dissipation strengths.  In the
Hamiltonian limit $\gamma = 0$, the current%
\footnote{Following Ref.~\cite{ac-quantum}, we used in the Hamiltonian
limit the initial condition $ \psi(x,t_{0}) = (2 \pi)^{-1/2}$,
\textit{i.e.},
the zero-momentum plane wave, and average the current over the
initial time $t_0$ of the driving.}
vanishes at the symmetry points $\theta = 0, \pi$ as discussed
above.  Moreover, it complies with relation (\ref{SymSym}).
For finite dissipation, the current exhibits multiple current
reversals upon changing the phase lag $\theta$.  This feature is
already very pronounced for $\gamma = 5\cdot 10^{-4}$, which
emphasizes that even very weak dissipation changes the behavior
significantly. For much stronger dissipation, $\gamma=0.02$, the
magnitude of the current changes slightly, while we still observe
similar current reversals.
For classical dissipative ratchets, such current reversals have been
attributed to tangent bifurcations when going from limit cycles
towards strange attractors \cite{Mateos}. In the Hamiltonian limit
of the quantum dynamics, however, these bifurcations are absent.
Therefore, the quantum current reversals may be attributed to
dissipation as well.

\begin{figure}[t]
\begin{center}
\includegraphics[width=0.95\linewidth,clip=true]{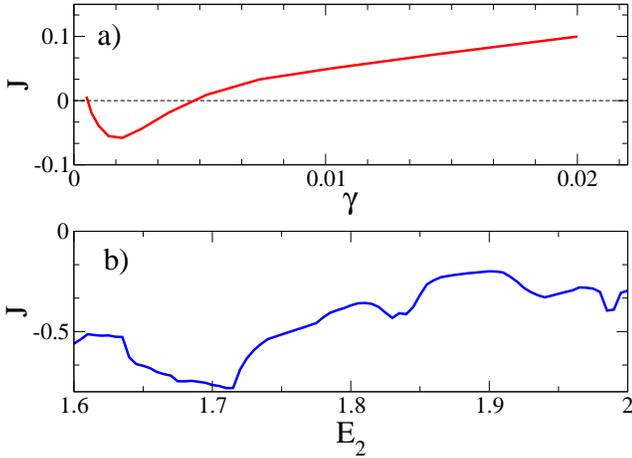}
\end{center}
\caption{(Color online) (a) Average current $J$ (in units of the
recoil momentum) as a function of (a) the dissipation strength
$\gamma$ ($E_2=2$, $\theta=0$) and (b)  as a function of the second
harmonic amplitude,  $E_2$ in (\ref{driv}) ($\gamma=0.02$,
$\theta=\pi/2$). The other parameters are the same as in Fig.3.}
\label{Fig:bifurc}
\end{figure}

We next focus on the symmetry point $\theta=0$, where the
Hamiltonian system (\ref{Ham2}) is time-reversal symmetric and
non-transporting. Time-reversal symmetry implies invariance under
$p\to -p$, which is perceivable in the Husimi representation of the
Floquet states at stroboscopic times shown in
fig.~\ref{Fig:Floquet}(a). Finite dissipation, however, destroys
time-reversal symmetry, such that the attractor looses the symmetry
$p\to -p$, see fig.~\ref{Fig:current}(b,c), despite the fact that it
is composed of symmetric, non-transporting Floquet states.  This
reveals that genuine quantum coherence, \textit{i.e.}\ off-diagonal density
matrix elements, play a crucial role for both the shape of the
attractor and the ratchet current.
Figure~\ref{Fig:bifurc}(a) shows the dependence of the ratchet current on the
dissipation strength $\gamma$. Both for $\gamma=0$ and in the limit
$\gamma\to 0$, the current vanishes.  For $\gamma>0$  we observe a
purely dissipation-induced quantum ratchet current.  This current is
negative for faint dissipation, but crosses zero and becomes
positive with increasing dissipation.  This current reversal
behavior resembles the one found for the corresponding classical
problem \cite{Bolt}, but even there has not been explained
analytically. The dependence of the ratchet current on the amplitude
of the second harmonic, $E_2$ in (\ref{driv}), is shown on
Fig.~\ref{Fig:bifurc}(b).

\section{Approaching the Hamiltonian limit}

For dissipative chaotic quantum systems, the limit of vanishing
dissipation, $\gamma\to 0$, deserves some attention.  This will reveal
an intriguing difference between classical and quantum ratchets.
Already in the presence of arbitrarily small dissipation, the
classical phase-space flow converges to a generally unique attractor,
while a Hamiltonian system in principle preserves memories to its
initial state for an arbitrarily long time.  Thus, for $\gamma=0$, the
classical current generally depends on the initial conditions, in
particular, if the initial phase-space distribution overlaps with
regular manifolds \cite{zaslavsky}.  If the initial condition lies
within a chaotic manifold, however, the system will in the long-time
limit be distributed uniformly over the chaotic manifold and, thus,
will eventually become independent of the initial condition \cite{Den,
Ketzmerick}.  For the corresponding quantum system, this ergodicity is
not found owing to the linearity of the Schr\"{o}dinger equation in
combination with the finite number of states involved.  As a
consequence, the limit of arbitrarily weak dissipation, $\gamma\to 0$,
may for quantum systems be different from the Hamiltonian case,
$\gamma=0$, as we observe in fig.~\ref{Fig:current}(a).


\section{Conclusions}

We have studied the quantum ratchet effect in the weakly dissipative
regime in which the quantum coherence suffers from
decoherence and relaxation. A central property of the corresponding,
unique quantum attractor is a quantum ratchet current, given by the
time-averaged momentum expectation value.  We found that even for
very weak dissipation, the current differs strongly from its
corresponding Hamiltonian counterpart.
The presence of any of two symmetries, namely generalized parity
and, in the Hamiltonian limit, time-reversal symmetry, inhibits a
ratchet current.  For bichromatic driving, the phase lag $\theta$
between the two harmonics determines whether the Hamiltonian part is
time-reversal symmetric or not. If time reversal holds, the current
vanishes in the Hamiltonian limit, while for finite dissipation, we
observe a purely dissipation-induced quantum ratchet current which,
moreover, possesses current reversals as a function of the
dissipation strength.

For cold atoms, the resulting currents are of the order 10--30\% of
the recoil momentum, being measurable with present experimental
techniques \cite{very_cold}.
Our study provides evidence that cold atoms in driven periodic
potentials are a natural candidate for studying the complex dynamics
originating from an intriguing interplay of nonlinearity, weak
quantum dissipation, and spatio-temporal symmetry violation.

\acknowledgments
This work has been supported by the DFG through
grant HA1517/31-1. Support by the German Excellence Initiative via
the ``Nanosystems Initiative Munich (NIM)'' is gratefully
acknowledged.


\begin{thebibliography}{10}

\bibitem{Smol}
\Name{Smoluchowski, M. v.} \REVIEW{Phys. Zeitschr.}{13}{1912}{1069}.

\bibitem{Han1}
\Name{Reimann P. \and H\"anggi P.} \REVIEW{Appl. Physics A}{75}{2002}{169};
\Name{H\"anggi P. \and Marchesoni F.} \REVIEW{Rev. Mod. Phys.}{}{in
press}{arXiv:0807.1283}

\bibitem{Reim}
\Name{Reimann P.} \REVIEW{Phys. Rep.}{361}{2002}{57}.

\bibitem{biolog}
\Name{J\"{u}licher F., Ajdari A. \and Prost J.}
\REVIEW{Rev. Mod. Phys.}{69}{1269}{1997}.

\bibitem{HH}
\Name{Astumian R. D. \and H\"anggi P.} \REVIEW{Phys. Today}{55 (11)}{2002}{33};
\Name{H\"{a}nggi P., Marchesoni F. \and Nori F.}
\REVIEW{Ann. Phys. (Leipzig)}{14}{2005}{51}.

\bibitem{Linke1999a}
\Name{Linke H. \textit{et al.}} \REVIEW{Science}{286}{1999}{2314}.

\bibitem{Khrapai2006a}
\Name{Khrapai V.~S. \textit{et al.}}
\REVIEW{Phys. Rev. Lett.}{97}{2006}{176803}.

\bibitem{Majer2003}
\Name{Majer J. B. \textit{et al.}} \REVIEW{Phys. Rev. Lett.}{90}{2003}{056802}.

\bibitem{Grifoni_spin}
\Name{Smirnov S.,  Bercioux D., Grifoni M. \and Richter K.}
\REVIEW{Phys. Rev. Lett.}{100}{2008}{230601}.

\bibitem{Flach1}
\Name{Flach S., Yevtushenko O. \and Zolotaryuk Y.}
\REVIEW{Phys. Rev. Lett.}{84}{2000}{2358}.

\bibitem{Den}
\Name{Denisov S. \textit{et al.}}
\REVIEW{Phys. Rev. E}{66}{2002}{041104}.

\bibitem{Kohler1}
\Name{Lehmann J., Kohler S., H\"{a}nggi P. \and Nitzan A.}
\REVIEW{J. Chem. Phys.}{118}{2003}{3283}.

\bibitem{ren}
\Name{Schiavoni M., Sanchez-Palencia L., Renzoni F. \and Grynberg G.,}
\REVIEW{Phys.  Rev. Lett.}{90}{2003}{094101};
\Name{Gommers R., Denisov S. \and Renzoni F.}
\REVIEW{Phys. Rev. Lett.}{96}{2006}{240604}.

\bibitem{very_cold}
\Name{Morsch O. \and Oberthaler M.} \REVIEW{Rev. Mod. Phys.}{78}{2006}{179}.

\bibitem{Ketzmerick}
\Name{Schanz H., Otto M.-F., Ketzmerick R. \and Dittrich T.}
\REVIEW{Phys. Rev. Lett.}{87}{2001}{070601};
\Name{Schanz H., Dittrich T. \and Ketzmerick R.}
\REVIEW{Phys. Rev.  E}{71}{2005}{026228}.

\bibitem{Casati0}
\Name{Carlo G. G. \textit{et al.}}
\REVIEW{Phys. Rev. A}{74}{2006}{033617}.

\bibitem{ac-quantum}
\Name{Denisov S., Morales-Molina L., Flach S. \and H\"{a}nggi P.}
\REVIEW{Phys. Rev. A}{75}{2007}{063424}.

\bibitem{Han_quantum} \Name{Reimann P., Grifoni M. \and H\"anggi P.}
\REVIEW{Phys. Rev. Lett.}{79}{1997}{10}.

\bibitem{Goychuk}
\Name{Goychuk I. \and H\"anggi P.} \REVIEW{Europhys. Lett.}{43}{1998}{503};
\Name{Goychuk I., Grifoni M. \and H\"anggi P.}
\REVIEW{Phys. Rev. Lett.}{81}{1998}{649};
\SAME{81}{1998}{2837}.

\bibitem{Grifoni_diss}
\Name{Grifoni M., Ferreira M. S., Peguiron J. \and Majer J. B.}
\REVIEW{Phys. Rev. Lett.}{89}{2002}{146801}.

\bibitem{Scheidl} \Name{Scheidl S. \and Vinokur V. M.}
\REVIEW{Phys. Rev. B}{65}{2002}{195305}.

\bibitem{Zueco}
\Name{Zueco D. \and Garcia-Palacios J. L.}
\REVIEW{Physica E}{29}{2005}{435}.

\bibitem{Han_Sm} \Name{Machura L. {\it et al.}}
\REVIEW{Phys. Rev. E}{70}{2004}{031107}.

\bibitem{Casati}
\Name{Carlo G. G., Benenti G., Casati G. \and Shepelyansky D. L.}
\REVIEW{Phys. Rev. Lett.}{94}{2005}{164101}.

\bibitem{Kohler} \Name{Kohler S., Dittrich N. \and H\"{a}nggi P.}
\REVIEW{Phys. Rev. E}{55}{1997}{300}.

\bibitem{Rotating} \Name{Grifoni M. \and  H\"{a}nggi P.}
\REVIEW{Phys. Rep.}{304}{1998}{232}.

\bibitem{amico}
\Name{Amico L., Osterloh A., \and Cataliotti F.} \REVIEW{Phys. Rev.
Lett.} {95}{2005}{063201}.

\bibitem{ring_exp}
\Name{Franke-Arnold S. \textit{et al}} \REVIEW{Opt. Exp.}
{15}{2007}{8619}.

\bibitem{Caldeira} \Name{Caldeira A. O. \and  Leggett A. L.}
\REVIEW{Ann. Phys. (N.Y.)}{149}{1983}{374}.

\bibitem{Weitz2007}
\Name{Salger T., Geckeler C., Kling S. \and Weitz M.} \REVIEW{Phys.
Rev. Lett.}{99}{2007}{190405}.

\bibitem{Husimi} \Name{Husimi K.}
\REVIEW{Proc. Phys. Math. Soc. Japan}{22}{1940}{264}.

\bibitem{Mateos}
\Name{Mateos J.} \REVIEW{Phys. Rev. Lett.} {84}{2000}{258}.

\bibitem{Bolt}
\Name{Yevtushenko O., Flach S., Zolotaryuk Y. \and Ovchinnikov A. A.}
\REVIEW{Europhys. Lett.} {54}{2001}{141}.

\bibitem{zaslavsky}
\Name{Zaslavsky G. M.}
  \Book{Physics of Chaos in Hamiltonian Systems}
  \Publ{Imperial College Press, London}
  \Year{1989}.

\end{thebibliography}
\end{document}